\documentclass[notitlepage,aps,twocolumn,superscriptaddress,longbibliography]{revtex4-1}
\usepackage{amsmath}
\usepackage{amssymb}
\usepackage{todonotes}
\usepackage[unicode=true,colorlinks=true,citecolor=blue,urlcolor=blue]{hyperref}

\usepackage{bm}
\usepackage{epsfig}
\usepackage{lipsum}

\usepackage[normalem]{ulem}

\renewcommand {\phi}{{\varphi}}

\begin{document}
\title{Non-trivial point-gap topology and non-Hermitian skin effect in photonic crystals}

\author{Janet Zhong}
\affiliation{Department of Applied Physics, Stanford University, Stanford, California 94305, USA}
\author{Kai Wang}
\affiliation{Department of Electrical Engineering, Ginzton Laboratory, Stanford University, Stanford, California 94305, USA}
\author{Yubin Park}
\affiliation{Department of Electrical Engineering, Ginzton Laboratory, Stanford University, Stanford, California 94305, USA}
\author{Viktar Asadchy}
\affiliation{Department of Electrical Engineering, Ginzton Laboratory, Stanford University, Stanford, California 94305, USA}
\author{Charles C. Wojcik}
\affiliation{Department of Electrical Engineering, Ginzton Laboratory, Stanford University, Stanford, California 94305, USA}
\author{Avik Dutt}
\affiliation{Department of Electrical Engineering, Ginzton Laboratory, Stanford University, Stanford, California 94305, USA}
\author{Shanhui Fan}
\email{shanhui@stanford.edu}
\affiliation{Department of Applied Physics, Stanford University, Stanford, California 94305, USA}
\affiliation{Department of Electrical Engineering, Ginzton Laboratory, Stanford University, Stanford, California 94305, USA}

\begin{abstract}
We show that two-dimensional non-Hermitian photonic crystals made of lossy material can exhibit non-trivial point gap topology in terms of topological winding in its complex frequency band structure. Such crystals can be either made of lossy isotropic media which is reciprocal, or lossy magneto-optical media which is non-reciprocal. We discuss the effects of reciprocity on the properties of the point gap topology. We also show that such a non-trivial point gap topology leads to non-Hermitian skin effect when the photonic crystal is truncated. In contrast to most previous studies on point gap topology which used tight-binding models, our work indicates that such non-Hermitian topology can studied in photonic crystals, which represent a more realistic system that has technological significance. 
\end{abstract}
\date{\today}

\maketitle

\section{Introduction} 
Topological band theory has played a central role in the development of condensed matter physics in the past two decades~\cite{bansil2016colloquium,kitaev2009periodic,hasan2019colloquium,bernevig2013topological,stanescu2017introduction}, and has received increased attention in other areas such as photonics and acoustics~\cite{ozawa2019topological,lu2014topological}. Topological band theory was initially developed for Hermitian systems. However, in recent years there have been growing interests in exploring topological band theory of non-Hermitian systems~\cite{kawabata2019symmetry,bergholtz2021exceptional,shen2018topological,torres2020perspective,midya2018nonhermitian,yokomizo2019nonbloch,gong2018topological,alvarez2018topological}. Such interests are in part motivated by the developments in topological photonics, where non-Hermitian features such as gain and loss are quite common and are essential for a number of device applications such as lasers and absorbers~\cite{feng2017nonhermitian,elganainy2019thedawn,harari2018topological,bandres2018topological}. \\

There are important connections as well as differences between the topological band theory of Hermitian and non-Hermitian systems~\cite{gong2018topological,bergholtz2021exceptional}. For topological classifications of Hermitian systems, the presence of gaps in the energy eigenvalue spectra plays an essential role~\cite{kitaev2009periodic}. To classify the topology of non-Hermitian systems, one needs to generalize the concept of the energy gap. Since in a non-Hermitian system the energy eigenvalues become complex, there are several different generalizations, including the point gap~\cite{gong2018topological,kawabata2019symmetry,bergholtz2021exceptional}, the line gap~\cite{kawabata2019symmetry,bergholtz2021exceptional}, and the separable band condition~\cite{wojcik2020homotopy,bergholtz2021exceptional}.\\ 

In this work, we focus on the point-gap topology.  An energy band is said to have a point gap at a reference energy if the reference energy does not lie on the energy band~\cite{gong2018topological}. For non-Hermitian systems, since the energy eigenvalues are complex, an energy band can form a loop in the complex energy plane. With respect to a reference energy located inside the loop, the energy band can thus acquire a non-zero winding number. Such a non-zero winding number defines a nontrivial point-gap topology with respect to the reference energy. We note that the point-gap topology is unique to non-Hermitian systems. The winding number as defined above is exactly zero for any Hermitian system. Also, the point-gap topology classifies the topology of the energy eigenvalues. This is in contrast to the topology of the Hermitian systems which classifies the topology of the energy eigenstates~\cite{bergholtz2021exceptional}. \\

The presence of nontrivial point-gap topology of an infinite bulk media manifests as non-Hermitian skin effect when the medium is truncated~\cite{yao2018edge,okuma2020topological,zhang2020correspondence,zhang2021universal,okuma2020topological,gong2018topological}. For a semi-infinite media with an open boundary condition on one edge, all states become localized on the edge and all of these states have energy lying inside the loop, with the degeneracy of the edge state equal to the winding number. The nontrivial point-gap topology also manifests in the spectrum of a finite system with open boundary condition on both edges~\cite{yao2018edge,gong2018topological}. \\

In this paper, we explore point-gap topology and non-Hermitian skin effects in photonic crystal systems. Previously these topological effects have been studied both theoretically and experimentally in various platforms such as optics~\cite{wang2021generating,weidemann2020topological,xiao2020nonhermitian}, electrical circuits~\cite{helbig2020generalized,liu2021nonhermitian}, mechanical systems~\cite{ghatak2020observation}, exciton-polaritons~\cite{xu2021interaction}, ultracold atoms~\cite{li2020topological} and acoustic crystals~\cite{zhang2021acoustic}. Exploring these effects in photonic crystals allows one to demonstrate non-Hermitian topology in a system that is of significant technological relevance~\cite{joannopoulos1997photonic}. Moreover, in all previous studies, the systems can be effectively modelled using tight-binding Hamiltonians, which describe the physics of the systems in terms of the coupling between discrete lattice sites that are close to each other. While the tight-binding approximation is valid for approximately localized atomic orbitals for electrons in solids, they are less applicable to media such as photonic crystals, where the interactions can be highly nonlocal~\cite{mongaleev2000long,dyerville2002tight,busch2003wannier}.   Thus, the study of effects related to point-gap topology in photonic crystals may bring new insights into non-Hermitian topological physics. Also related to our work here, complex band structures in non-Hermitian photonic crystals have been well studied, particularly in work focusing on $\mathcal{PT}$-symmetry, exceptional points and unidirectional transmission~\cite{ding2015coalescence,kim2006exact,fietz2011complex}. There have also been various studies on non-Hermitian topological band structure in photonic crystals~\cite{zhou2020topological,prudencio2020first} but these studies focus only on line-gap topology. To the best of our knowledge, no previous studies have investigated non-Hermitian topological photonic crystals from a point-gap topology perspective. \\

The rest of the paper is organized as follows: In Sec.~\ref{sec_theory}, we describe the theoretical condition for achieving nontrivial point-gap topology in photonic crystals. We show that in one-dimensional systems nonreciprocity is essential for achieving nontrivial point-gap topology. In contrast, in two-dimensional systems, it is possible to achieve nontrivial point-gap topology along various loops inside the first Brillouin zone in a reciprocal system provided that certain spatial symmetry is broken. Motivated by the theoretical conditions as discussed in Sec.~\ref{sec_theory}, in Sec.~\ref{numerical}, we present numerical studies of nontrivial point-gap topology and non-Hermitian skin effects in both reciprocal and nonreciprocal photonic crystal structures. We conclude in Sec.~\ref{conclusion}.

\section{Theoretical background}
\label{sec_theory}
\begin{figure}[h]
\centering
\includegraphics[width=0.45\textwidth]{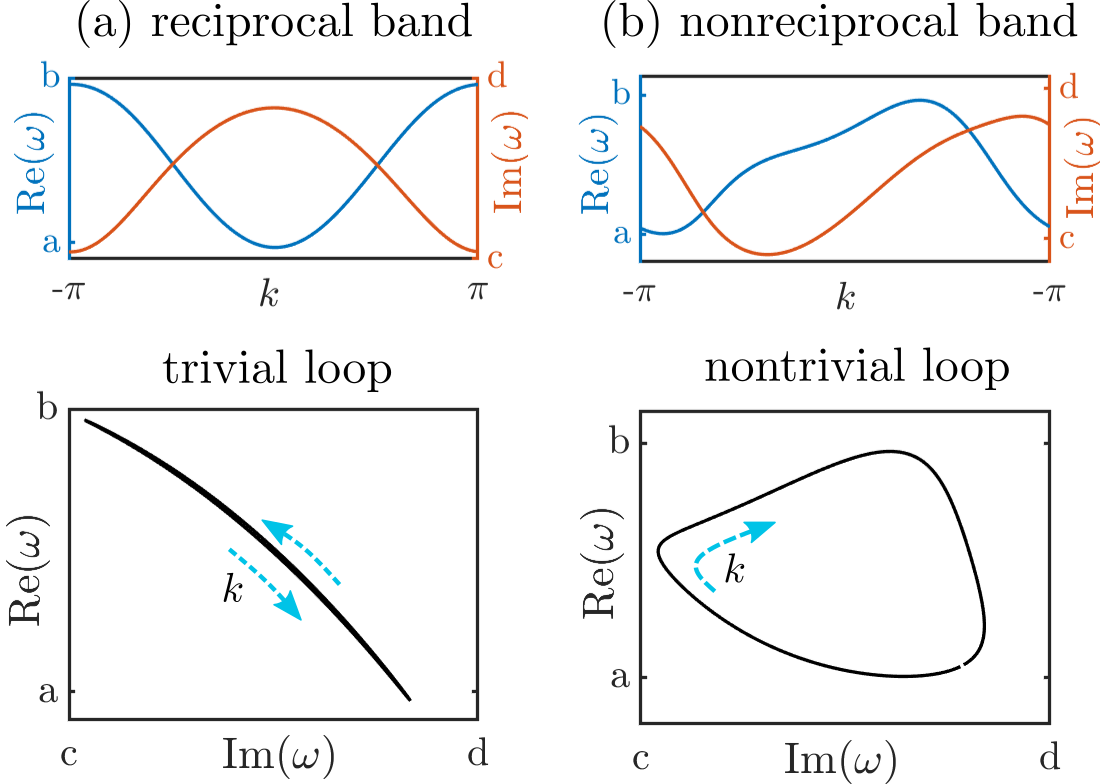}
\caption{(a) Reciprocal complex band structures in 1D lead to trivial windings while (b) nonreciprocal complex bands can give rise to nontrivial windings.}
\label{loops}
\end{figure}
The non-Hermitian point-gap topology for a single energy band in a 1D system corresponds to a closed loop traced by the band structure $\omega(k)$ in the complex frequency plane. The point-gap topology is characterised by the winding number defined as:
\begin{equation}
W:=\int_{-\pi/a}^{\pi/a} \frac{\mathrm{d} k}{2 \pi i} \frac{\mathrm{d}}{\mathrm{d} k} \ln [\omega(k)-\omega_0] \:,
\label{1dwinding}
\end{equation}
where $\omega$ is a normalized complex eigenfrequency, $k \in [-\pi/a, \pi/a$) is a wavevector in 1D with $a$ being the periodicity and $\omega_0 \in \mathbb{C}$ is a reference angular frequency. Here, $W$ gives the number of times the complex eigenfrequency winds around $\omega_0$ as $k$ varies across the first Brillouin zone. The direction of winding gives the sign of $W$, with the clockwise (anticlockwise) direction corresponding to $W=-1$ ($W=1$). For simplicity, in this paper we will only consider complex loops made by a single band. However, the complex loops could also in principle be made of multiple bands for photonic crystals. In this case, one would replace $\omega(k)$ with $\det (H(k)- \omega_0)$ in Eq.~\eqref{1dwinding} where $H(k)$ is an $M \times M$ matrix block associated with the $M$ bands of interest in the multi-band Hamiltonian in $k$-representation~\cite{gong2018topological}. \\

In order to achieve a nontrivial winding in a 1D system, a necessary condition is to ensure that $\omega(k) \neq \omega(-k)$ (Fig.~\ref{loops}(b)). If $\omega(k)= \omega(-k)$ for all $k$, the band structure always has trivial winding (a loop with no enclosed area), as the band necessarily retraces itself in the complex plane as $k$ varies from one end of the Brillouin zone to the other (Fig.~\ref{loops}(a)). Thus, the system cannot have any spatial symmetry that maps between the states at $k$ and $-k$, such as spatial inversion or mirror symmetry. Moreover, the system must break reciprocity, since in a reciprocal system $\omega(k)=\omega(-k)$~\cite{figotin2001nonreciprocal}. (We include a proof of this fact for photonic crystals in the \hyperref[appendix]{Appendix}). \\

The winding number in Eq.~\eqref{1dwinding} is defined for a 1D system. For a 2D system with a band structure defined by $\omega\left(k_{x}, k_{y}\right)$, we can assign a similar winding number for any loop inside the first Brillouin zone. As a simple illustration, in this paper, we only consider a square lattice with a lattice constant $a$. At any given $k_{y}$, the straight path from $\left(k_{x}=-\frac{\pi}{a}, k_{y}\right)$ to $\left(k_{x}=\right.$ $\frac{\pi}{a}, k_{y}$) forms a closed loop due to the periodic property of the first Brillouin zone. Therefore, one can define a $k_{y}$ -dependent winding number
\begin{equation}
W\left(k_{y}\right)=\int_{-\pi/a}^{\pi/a} \frac{\mathrm{d} k_x}{2 \pi i} \frac{\mathrm{d}}{\mathrm{d} k_x} \ln [\omega(k_x,k_y)-\omega_0] 
\end{equation}
as a straightforward generalization of the winding number for 1D systems. Likewise, one can also define a $k_x$-dependent winding number by tracing out $k_y.$\\

Similar to 1D systems, in order to achieve a non-zero winding number $W\left(k_{y}\right)$ for a given $k_{y}$, the structure cannot have any symmetry that maps between the states at $\left(k_{x}, k_{y}\right)$ and $\left(-k_{x}, k_{y}\right)$, which implies a necessity to break mirror symmetry with respect to the $k_x=0$ axis. Therefore, the photonic crystal cannot have a mirror plane perpendicular to the $x$ direction in real space as this would imply mirror symmetry about the $k_x=0$ axis. For reciprocal systems, as proved in the \hyperref[appendix]{Appendix}, the band structure in addition satisfies
\begin{equation}
\omega\left(k_{x}, k_{y}\right)=\omega\left(-k_{x},-k_{y}\right),
\label{wkxky}
\end{equation}
Therefore, if $k_{y}$ and $-k_{y}$ differ by a reciprocal lattice vector, $W\left(k_{y}\right)$ has to vanish. Thus, for the square lattice case, by reciprocity we must also have
\begin{equation}
W\left(k_{y}=0\right)=W\left(k_{y}=\frac{\pi}{a}\right)=0.
\label{wkxky_0}
\end{equation}
At $k_y = 0$ obviously $k_y = - k_y$. At $k_y = \pi/a$, we note that $-k_y = -\pi/a$ differ from $\pi/a$ by a reciprocal lattice vector of $2\pi/a$ $\hat{y}$, where $\hat{y}$ is the unit vector along the $y$ direction. On the other hand, away from these special $k_{y}$ points, $W\left(k_{y}\right)$ can be nonzero even for a reciprocal system. Thus, in a reciprocal 2D system, it is possible to achieve nontrivial winding. Moreover, Eq.~\eqref{wkxky} implies that
\begin{equation}
W\left(k_{y}\right)=-W\left(-k_{y}\right) \:.
\label{wkyneg}
\end{equation}

In Sec.~\ref{secrec}, we will show an example of a reciprocal photonic crystal design with nontrivial $W(k_y)$. For a nonreciprocal 2D system, on the other hand, the constraint of Eq.~\eqref{wkxky} is no longer applicable. Thus, provided that the relevant spatial symmetries are also broken, we can have nonzero $W(k_y)$ at $k_y=0$ and $\pm \frac{\pi}{a}$ and we no longer require $W\left(k_{y}\right)=-W\left(-k_{y}\right)$. We will give an example of such a nonreciprocal photonic crystal structure in Sec.~\ref{secnon}. \\

As was noted in Ref.~\cite{gong2018topological}, for a one-dimensional system, the winding number of the bulk band structure is directly connected to the behavior of the edge states when the structure is truncated. In particular, for a semi-infinite system with an edge on the left (right) side, there are localized edge states at every complex energy inside the loop with a positive (negative) winding number. The degeneracy at each of these complex energy is equal to the winding number. For finite systems truncated with open boundary conditions on both edges, the energy spectrum of the edge states in general forms lines inside the loop. These edge-state behaviors are commonly referred to as the non-Hermitian skin effect~\cite{yao2018edge}. For the two-dimensional system as considered here, at each fixed $k_y$ the system is one dimensional. Therefore, to probe non-Hermitian skin effects, we can consider a stripe geometry, in which the system is periodic along the $y$-direction and truncated along the $x$-direction. With such a stripe geometry, the winding number behaviors discussed above should manifest in the non-Hermitian skin effects of the corresponding semi-infinite and finite systems. \\

Based on the theoretical discussion above, in the next section we present numerical results for both reciprocal and nonreciprocal photonic crystal structures. We consider both the point-gap topology of the bulk bands, as well as the corresponding edge state behaviors.\\

\section{Numerical results}
\label{numerical}
\subsection{Reciprocal 2D photonic crystal}
\label{secrec}
\begin{figure}[h]
\centering
\includegraphics[width=0.5\textwidth]{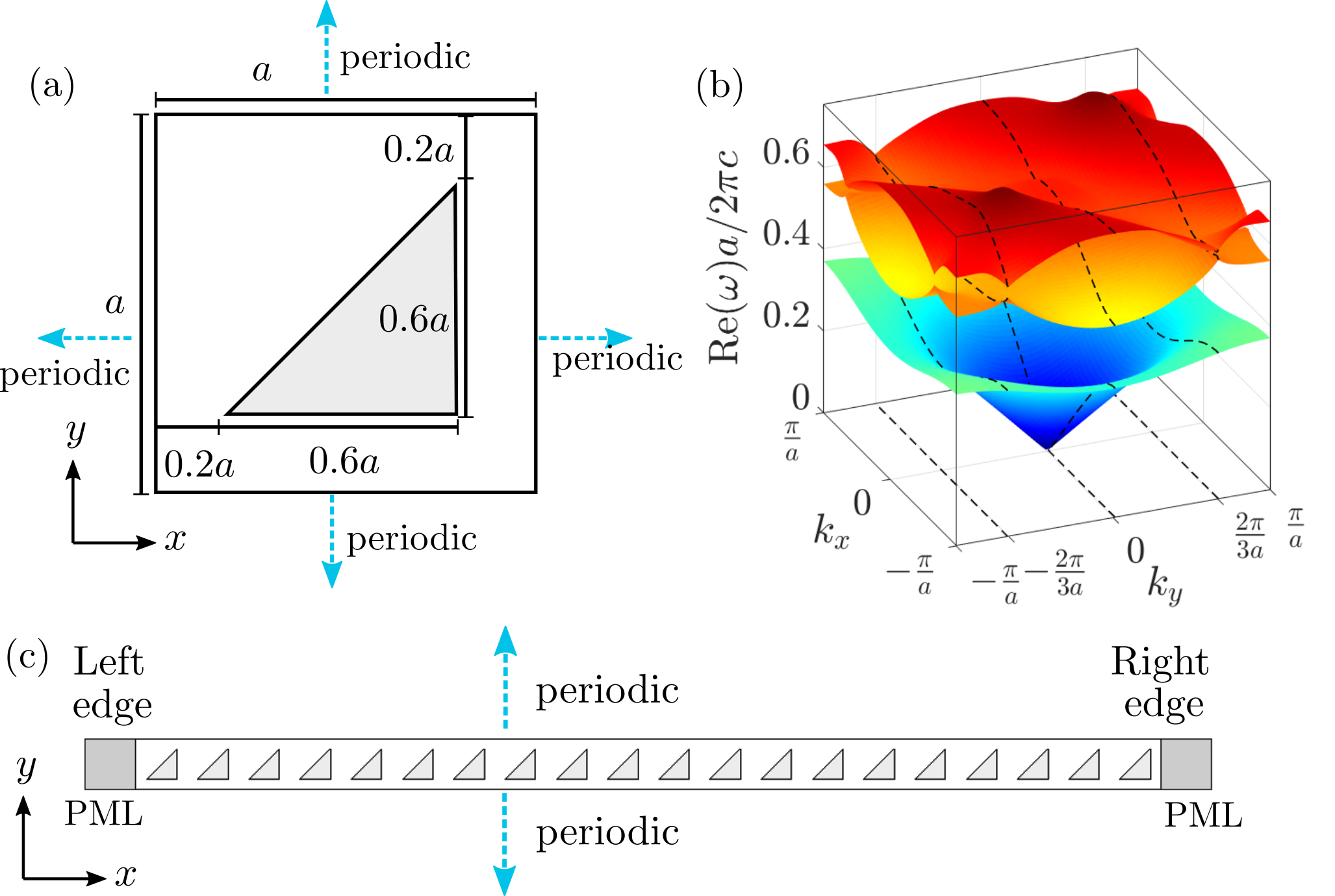}
\caption{(a) Unit cell of photonic crystal where the lattice periodicity is $a$. Here, the background is air and the triangle has refractive index $n_{\text{tri}}=2-1i$. (b) The lowest three bands of the real part of the  eigenfrequences in the first Brillouin zone for the photonic crystal geometry in (a). The dashed black lines at fixed $k_y=\frac{-2\pi}{3a},0$ and $\frac{2\pi}{3a}$ indicate the path traced for the reduced 1D $k_x$-space band diagrams plotted in Fig.~\ref{skin}. (c) Finite stripe geometry that is periodic along $y$ and has finite $N=20$ unit cells in the $x$ direction. The ends in the $x$-direction have perfectly matched layers (PML) that are matched to air.}
\label{setup}
\end{figure}

\begin{figure*}[t!]
\centering\includegraphics[width=\textwidth]{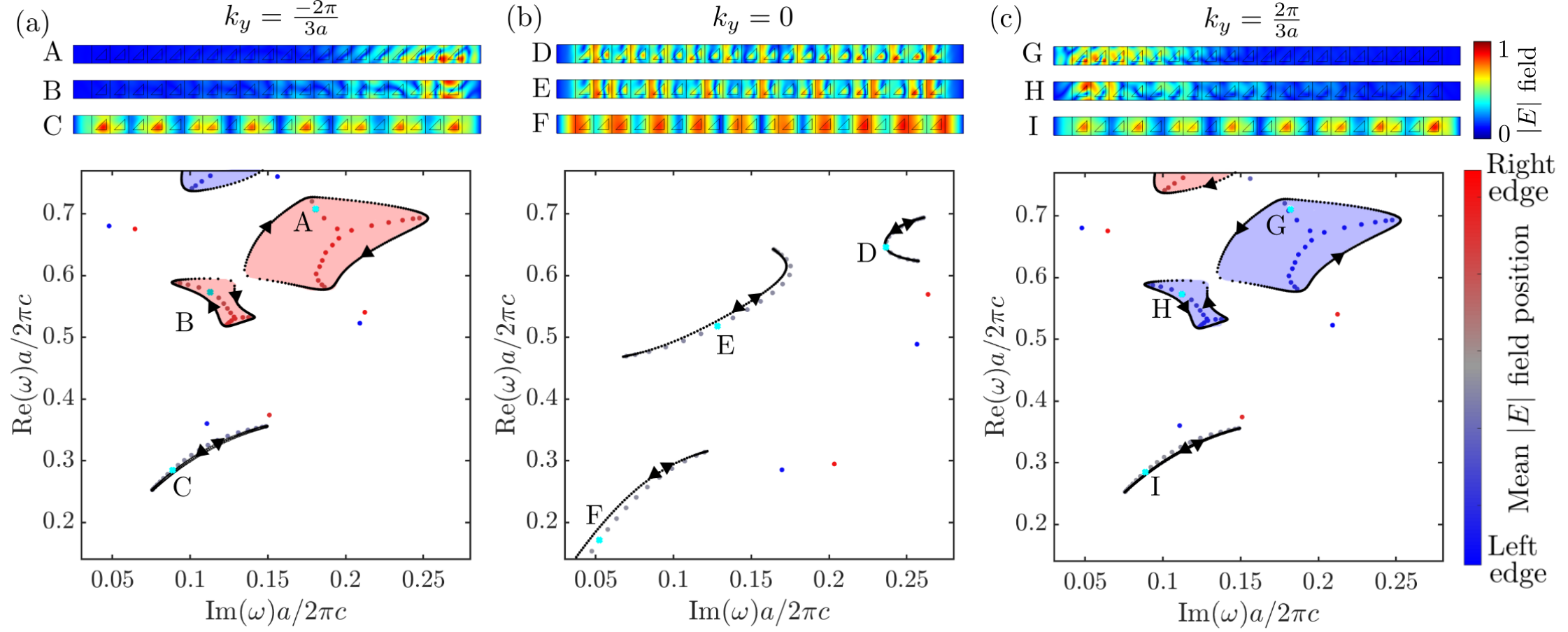}
\caption{The lower part of each panel shows the band structure of the infinite 2D photonic crystal, plotted on the complex frequency plane, as well as the spectra of the finite stripe geometry. The band structure is plotted in black and the arrows give the direction of winding as $k_x$ varies parametrically from $k_x = -\pi/a$ to $\pi/a$. Clockwise winding corresponds to $W=-1$ and the enclosed area is shaded red whereas anti-clockwise winding corresponds to $W=1$ and the enclosed area is shaded blue. The eigenfrequencies of the finite stripe geometry is colored by mean $|E(x,y)|$ field position, where red signifies an electric field distribution localized at the right edge and blue signifies electric field localization localized at the left edge. The upper panels represent the electric field distribution of the eigenmodes of the stripe geometry at various complex frequencies indicated in the bottom panel. (a), (b) and (c) correspond to $k_y=\frac{-2\pi}{3a},0$ and $k_y=\frac{2\pi}{3a}$ respectively and A-I are the normalized electric field eigenstates. }
\label{skin}
\end{figure*}
In this section, we consider a 2D square-lattice photonic crystal with lattice periodicity $a$. Each unit cell (Fig.~\ref{setup}(a)) consists of an isosceles right-angle triangle where the two legs have length 0.6$a$. The geometry of the unit cell is chosen such that both mirror symmetries in the $x$ and $y$ directions are broken. The triangle region consists of dielectric material with refractive index $n = 2-i$ and is surrounded by air ($n = 1$). Throughout the paper, we use the $e^{i\omega t}$ convention. Thus, a negative imaginary refractive index component means that the material inside the triangle is lossy and the structure is non-Hermitian.  \\

We numerically determine the photonic band structure of the photonic crystal system shown in Fig.~\ref{setup}(a) using COMSOL Multiphysics, which employs finite-element methods to numerically solve Maxwell's equations. Here, we consider TE waves which have their magnetic field in the $xy$-plane. Since the material in the triangle is lossy for a given $\mathbf{k}=(k_x,k_y)$, the eigenvalues of the energy are in general complex. The real part of the frequency eigenvalues Re($\omega$) as a function of wavevector $\mathbf{k}$ in the first Brillouin zone is plotted in Fig.~\ref{setup}(b) for the three photonic bands with lowest Re($\omega$). Where it is unambiguous in the plot, we will refer to the band with the lowest Re($\omega$) band as the first band, the second lowest as the second band, etc. \\

In the bottom panels of Fig.~\ref{skin}, we plot the same band structure $\omega\left(k_{x}, k_{y}\right)$ as shown in Fig.~\ref{setup}(b), along various lines in the first Brillouin zone at fixed $k_{y}$, with $k_{y}=0, \pm \frac{2 \pi}{3 a}$. At each fixed $k_{y}$, we vary $k_{x}$ from $-\frac{\pi}{a}$ to $\frac{\pi}{a}$ and plot the corresponding complex frequency eigenvalues in the complex frequency plane. Since $k_{x}=-\frac{\pi}{a}$ and $k_{x}=\frac{\pi}{a}$ are equivalent, the frequencies of the bands form loops in the complex frequency plane. At $k_{y}=0$, all bands exhibit trivial point-gap topology, which is consistent with Eq.~\eqref{wkxky_0}. As $k_{x}$ varies, the frequency in each band traces along a curve segment and eventually returns to the starting point. The resulting loop does not enclose any interior points. At $k_{y}=\pm \frac{2 \pi}{3 a}$, the second and third bands exhibit nontrivial point-gap topology. Each forms a loop that encloses an area in the complex frequency plane. The frequency eigenvalues wind nontrivially around any frequency value within the area. Such nontrivial winding is consistent with the theoretical discussion above. Also, for the same band the directions of winding change sign between $k_{y}=\pm \frac{2 \pi}{3 a}$. This is a consequence of reciprocity as noted in Eq.~\eqref{wkyneg}. Thus, the results here provide an illustration of the theory as discussed in the previous section. We also note that the first band does not seem to exhibit any nontrivial point-gap topology at $k_{y}=\pm \frac{2 \pi}{3 a}$. This may be because for the first band, the structure behaves as an effective uniform medium since the corresponding wavelength is significantly larger than the periodicity for most of the first Brillouin zone.\\

In the band theory for one-dimensional non-Hermitian systems, nontrivial point-gap topology of bulk system manifests in the edge state behavior as manifested by the non-Hermitian skin effect. In a corresponding finite system truncated by open boundary conditions, all eigenstates are localized at the edge. Moreover, the eigenfrequency spectra of the edge states fall within the areas enclosed by the bulk bands~\cite{gong2018topological,yao2018edge,bergholtz2021exceptional}. The non-Hermitian skin effects however are typically studied within tight-binding models. Here we show that such non-Hermitian skin effects manifest in the two-dimensional photonic crystal systems. For this purpose, we consider a structure shown in Fig.~\ref{setup}(c), which is periodic along the $y$-direction, and contains 20 unit cells in the $x$-direction. In the $x$-direction, outside the structure we include perfectly matched layers~\cite{berenger2007perfectly} (PML) that are matched to air. We have chosen to use PML at the boundaries as PML allows us to simulate air outside the photonic crystal, which is more realistic. We note that this boundary condition imposed on the edges of the photonic crystal is not the same as the open boundary condition commonly used in the tight-binding model~\cite{busch1987tight}. \\

Our finite stripe geometry is still periodic along $y$ and so we can plot the eigenvalues corresponding to specific $k_y$. In the bottom panels of Fig.~\ref{skin}, we plot the eigenfrequencies of such a finite stripe for $k_y = \frac{-2\pi}{3a}, 0$ and $\frac{2\pi}{3a}$ as colored dots. Here, the color represents the mean position $\bar{x}$ of the electric field norm $|E(x,y)|$ defined as 
\begin{equation}
\bar{x}=\frac{\int_0^{a}dy\int_0^{Na}dx |E(x,y)| x  }{ \int_0^{a}dy\int_0^{Na}dx |E(x,y)|}.
\end{equation}
where $N$ is the number of unit cells in the $x$ direction. The colorscale is chosen such that eigenstates with electric field localized at the left edge of the stripe ($x=0$) are colored blue, and eigenstates with field localized at the right edge ($x=20a$) are colored red. Delocalized eigenstates have a fairly central $\bar{x}$ and are colored grey. In panels A-I, we plot the corresponding eigenstates of the electric field norm $|E|$ corresponding to various eigenfrequencies labelled in the bottom panels of Fig.~\ref{skin}. \\

For the bands that exhibit trivial point-gap topology, (i.e. all three bands at $k_{y}=0$, as well as the first bands at $k_{y}=\pm \frac{2 \pi}{3 a}$), the corresponding eigenfrequencies of the finite stripe overlap with the bands of the infinite structure with only slight differences due to finite size effects of the stripe geometry. The overlap between the finite stripe and the bands becomes closer with increasing number of unit cells, $N$. The eigenstates corresponding to trivial bands (panel $\mathrm{C}$ for $k_{y}=-\frac{2 \pi}{3 a}$, panels $\mathrm{D}, \mathrm{E}, \mathrm{F}$ for $k_{y}=0$, and panel $\mathrm{I}$ for $k_{y}=\frac{2 \pi}{3 a}$) are delocalized inside the stripe. For the bands that exhibit nontrivial point-gap topology (i.e. the top two bands at $k_{y}=\pm \frac{2 \pi}{3 a}$), the corresponding eigenfrequencies of the finite stripe lie in the areas enclosed by the bands. The corresponding eigenstates (panel $\mathrm{A}$ and $\mathrm{B}$ for $k_{y}=-\frac{2 \pi}{3 a}$, panels $\mathrm{G}$ and $\mathrm{H}$ for $k_{y}=\frac{2 \pi}{3 a}$) are localized on the edges. Moreover, for these two $k_{y}$ values, the eigenstates of the stripe are localized at opposite edges, in consistency with the theoretical studies of non-Hermitian skin effects which correlate the edge where the state is localized with the direction of the winding of a bulk band~\cite{gong2018topological}. For our setup, a winding number of $-1$ ($1$) corresponds to localisation on the right (left) edge of the finite stripe. The numerical results here indeed indicate that the non-Hermitian skin effect can be studied in the photonic crystal system, in spite of the use of a typical interface between photonic crystal and air which is not the same as the open boundary condition assumed in many previous theoretical tight-binding studies. \\

For the stripe structure, in addition to the states associated with the bulk bands, there are also additional states with frequencies significantly outside either the bulk band or the encircled regions associated with them. These appear to occur as pairs in Fig.~\ref{skin}(a-c, bottom), where one is localized at the left (colored blue) and the other is localized at the right (colored red). These states are the edge states associated with the presence of the line gap and in our system may or may not have a topological origin. The number of these states does not increase as the number of unit cell increases and the frequencies of these states are located at isolated points. In contrast, the number of the localized edge states associated with the bulk band, as arised from the non-Hermitian skin effects, does increase as the number of unit cell increases. And the frequencies of these states form essentially continuous lines in the complex frequency plane in the limit of large number of unit cells.  \\

\begin{figure}[h]
\centering\includegraphics[width=0.4 \textwidth]{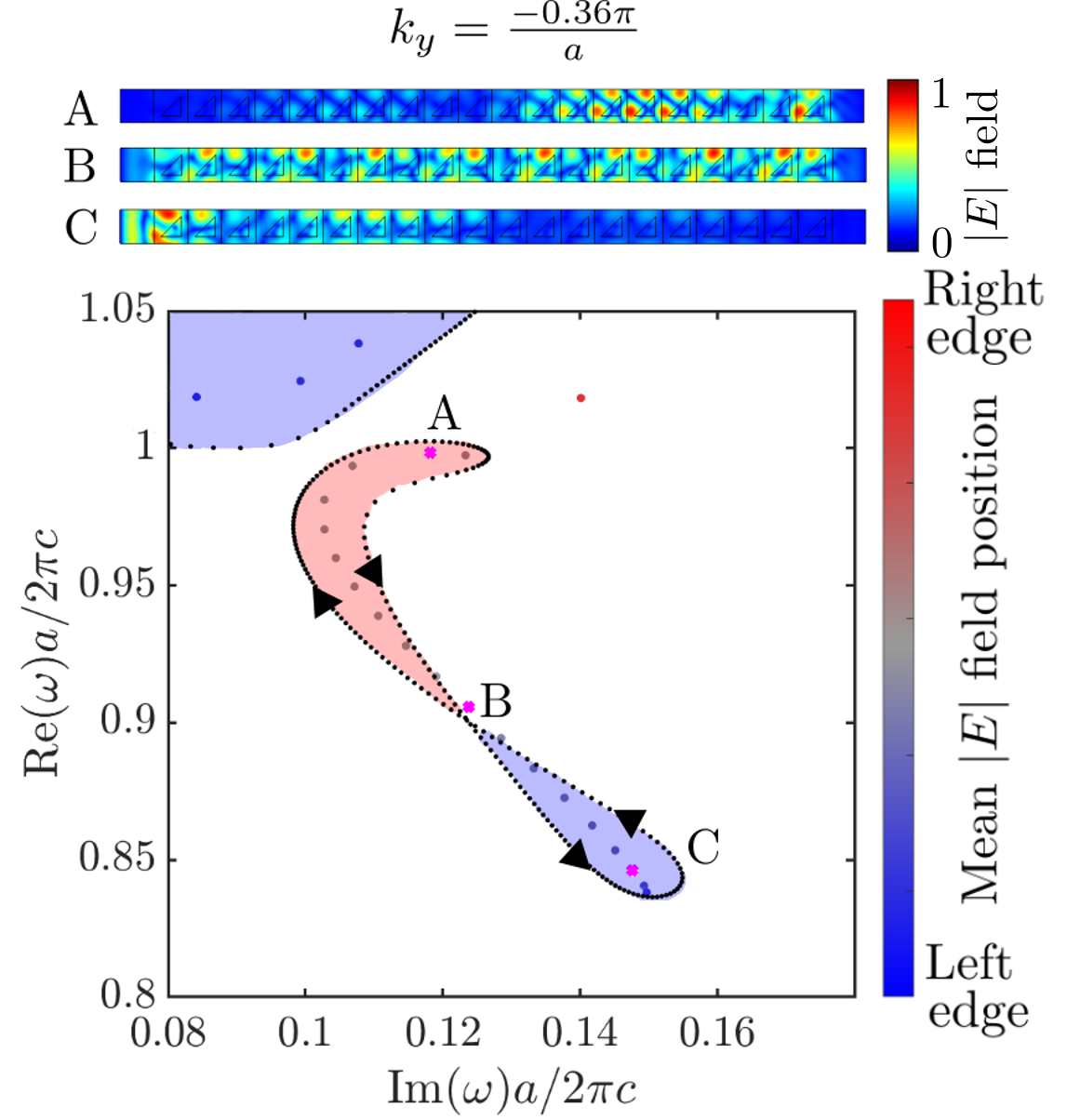}
\caption{Same as Fig.~\ref{skin}, except with $k_y = -0.36\pi/a$, and at a range of real part of frequencies that are higher as compared with Fig.~\ref{skin}(a). Point B is the Bloch point.}
\label{figureeight}
\end{figure}
In our system, we can also find more complicated winding patterns, such as a figure eight (or twisted loop~\cite{zhang2021acoustic}) winding, depicted in Fig.~\ref{figureeight}. Here, we use the same parameters as Fig.~\ref{skin} but consider the system at $k_y=\frac{-0.36 \pi}{a}$. A bulk band exhibits a loop that winds around two separate regions with opposite winding directions. For the finite stripe structure as shown in Fig.~\ref{setup}(c), the eigenstates associated with this band have their frequencies forming in a line in these two regions. The eigenstates in these two regions are localized on opposite edges as shown in Fig.~\ref{figureeight}, panel A and C, in consistency with the correlations between the winding direction of the bulk band and the localization behavior as commented previously. The bulk band self intersects at the “Bloch point”~\cite{song2019non, longhi2019probing} (Point B in Fig.~\ref{figureeight}). Corresponding to the Bloch point of the bulk band, the finite stripe has an eigenstate that is delocalized and with its eigenfrequency very close to the Bloch point, as shown in panel B of Fig.~\ref{figureeight}. Bloch points are interesting because they signify an unusual topological phase transition between two class of edge states localized on the opposite ends of the stripe. For this reason, the Bloch point has been said to correspond to a `bipolar skin effect~\cite{song2019non}.' Delocalized eigenstates at discrete points are unique to non-Hermitian systems and may have applications such as in extended laser modes~\cite{song2019non}. Bloch points were recently experimentally demonstrated in topological acoustic crystals for a tight-binding system~\cite{zhang2021acoustic}. Here we show that they can be realized in reciprocal photonic crystal structures. \\

\begin{figure}[h]
\centering\includegraphics[width=0.4 \textwidth]{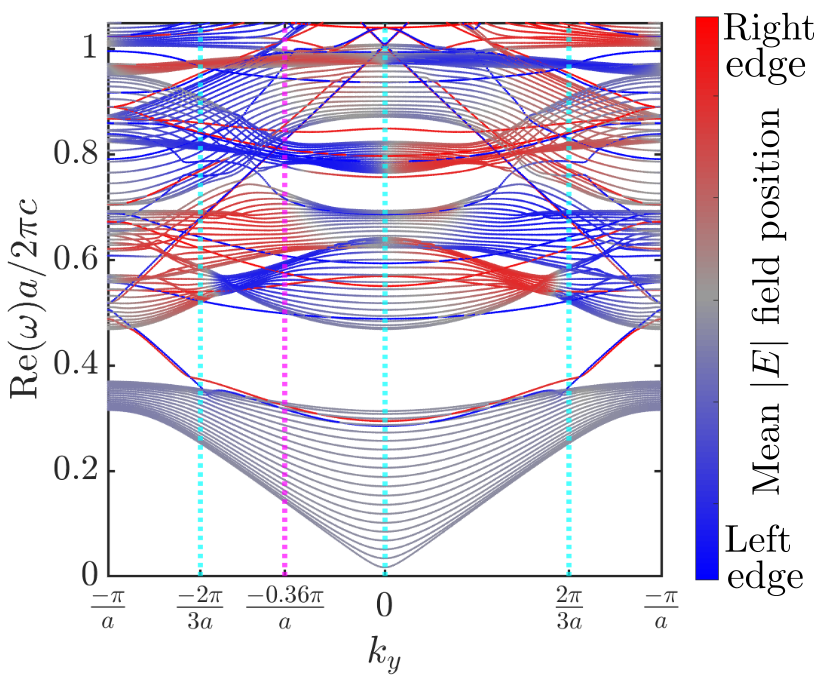}
\caption{Real part of the eigenfrequencies as a function of $k_y$ for the finite stripe geometry shown in Fig.~\ref{setup}(c), where the triangle consists of a reciprocal material. The color indicates the mean position of the electric field for each eigenmodes. }
\label{colouredbands}
\end{figure}
In Fig.~\ref{colouredbands}, we represent the bandstructure of the stripe geometry by plotting the real part of the eigenfrequencies as a function of $k_y$ for the stripe geometry shown in Fig.~\ref{setup}(c). We use the same color scheme as used in Fig.~\ref{skin} to represent the mean position $\bar{x}$ of the eigenstate. The specific $k_y$ values that are considered in Figs.~\ref{skin} and~\ref{figureeight} are denoted using vertical lines. We note that with the exception of the first band, significant part of all the upper bands are localized on either edges, in consistency with the results that we show in Figs.~\ref{skin} and~\ref{figureeight} for the cases of specific $k_y$ values. Also, the bandstructure is symmetric around $k_y = 0.$ And two states at $k_y$ and $-k_y$ are localized at opposite edges, in consistency with the theoretical discussions in Sec.~\ref{sec_theory}.  \\

To summarize this section, we show that both the nontrivial point-gap topology, and the non-Hermitian skin effects, can be seen in a two dimensional photonic crystal made of reciprocal lossy materials.

\subsection{Nonreciprocal 2D photonic crystal}
\label{secnon}
In the previous section we have discussed aspects of non-Hermitian topology in a reciprocal two-dimensional photonic crystal. In this section, to highlight the impact of reciprocity on non-Hermitian topology, we consider a 2D photonic crystal with the same geometry as Fig.~\ref{setup}(a) but where the triangle structure is made of magneto-optical material. We choose the magnetization direction of the magneto-optical material to be parallel to the $z$ axis, and the relative permittivity tensor is then~\cite[Sec.~2.2]{zvezdin1997modern}:
\begin{equation}
\varepsilon=\left[\begin{array}{ccc}
\varepsilon_d & -i \varepsilon_a & 0 \\
i \varepsilon_a & \varepsilon_d & 0 \\
0 & 0 & \varepsilon_{z} 
\end{array}\right]\:.
\label{eq_mag}
\end{equation}
In Eq.~\eqref{eq_mag}, $\varepsilon_a$ controls the strength of the nonreciprocity as resulting from the magnetization of the material. Here, we choose $\varepsilon_d=4.2-0.6i$ and $\varepsilon_a=5$. We consider a TM-polarized wave propagating in the $xy$-plane with magnetic field in the $z$ direction and electric field in the $xy$-plane. For such a wave, the value of $\varepsilon_z$ does not affect its properties. Nonreciprocity of such a strength, as measured by the ratio $\varepsilon_a/\varepsilon_d$, can be achieved in magnetic Weyl semimetals in the mid-infrared~\cite{kotov2018giant}. Alternatively, similar strength of nonreciprocity can be realized in the magnetic permeability tensor of ferrite materials in the microwave frequency range~\cite[Sec. 9.1]{pozar2011microwave}. To utilize such nonreciprocity one can consider the TE polarization for a 2D photonic crystal~\cite{wang2009observation,asadchy2015functional}. \\

\begin{figure}[h]
\centering
\includegraphics[width=0.40\textwidth]{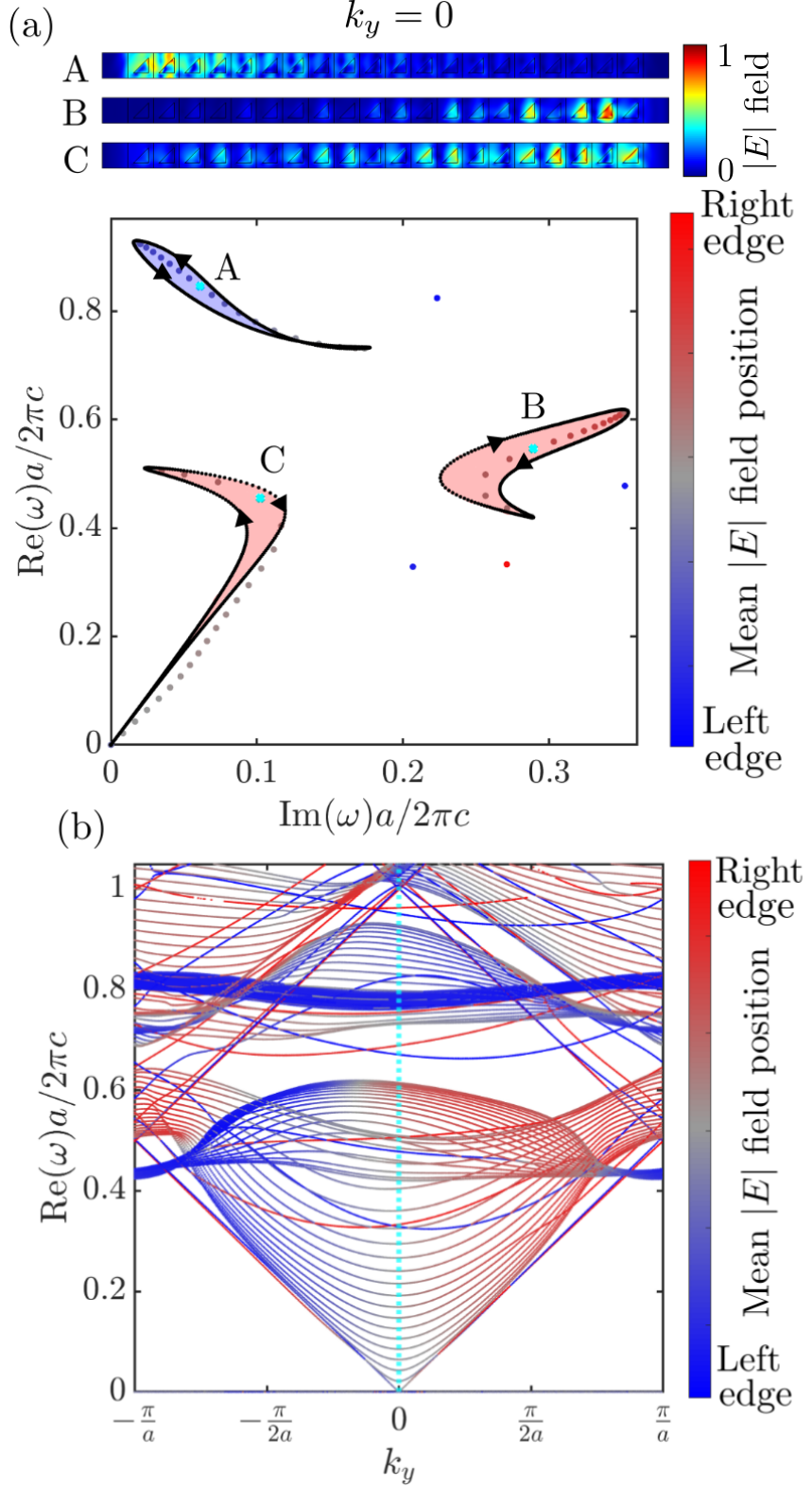}
\caption{(a) Same as Fig.~\ref{skin}(b), except with a non-reciprocal photonic crystal. (b) Real part of the eigenfrequencies as a function of $k_y$ for the finite stripe geometry shown in Fig.~\ref{setup}(c), where the triangle consists of a non-reciprocal material.  The color indicates the mean position of the electric field for each eigenmodes.}
\label{magneto}
\end{figure}

In the bottom panel of Fig.~\ref{magneto}(a), we plot the bandstructure for this nonreciprocal case for fixed $k_y=0$ as we vary $k_x$ from $-\frac{\pi}{a}$ to $\frac{\pi}{a}$. Unlike the reciprocal case in Fig.~\ref{skin}(c), here at $k_y = 0$ the band forms loops and exhibits nontrivial point-gap topology. The first two bands have $W=-1$ and the third band has $W=1$. Nontrivial loops can be achieved because nonreciprocity breaks the condition in Eq.~\eqref{wkxky} and so Eq.~\eqref{wkxky_0} no longer applies. In the bottom panel of Fig.~\ref{magneto}(a), we also plot the eigenvalues of the stripe geometry in Fig.~\ref{setup}(c) for the nonreciprocal case as coloured dots using the same color code as Fig.~\ref{skin}. Eigenstates corresponding to negative winding are localized at the right (panel B and C) and eigenstates corresponding to positive winding are localized at the left (panel A) which is consistent with the correspondence found in the reciprocal case. \\

In Fig.~\ref{magneto}(b), we plot the bandstructure of the stripe geometry for the nonreciprocal case by plotting the real part of the eigenfrequencies as a function of $k_y$, where we again use the same color code as Fig.~\ref{skin}. Here, $k_y=0$ (Fig.~\ref{magneto}(a)) is denoted using a vertical line. We note that unlike the reciprocal case in Fig.~\ref{colouredbands}, the bandstructure is not symmetric about $k_y=0$ and we can also have nontrivial loops at $k_y=0, \pm \frac{\pi}{a}$, which shows that Eq.~\eqref{wkxky} to Eq.~\eqref{wkyneg} do not apply for nonreciprocal media as expected.

\section{Conclusion}
\label{conclusion}
In summary, we have shown that various aspects of non-Hermitian point-gap topology, including band winding in complex frequency plane, and the non-Hermitian skin effect, can be realized in two-dimensional lossy photonic crystals that are either reciprocal or non-reciprocal. We also demonstrate some of the qualitative differences in non-Hermitian topology between Hermitian and non-Hermitian systems. While our calculations are for two-dimensional systems, we expect that similar effects should be seen in photonic crystal slab system where light experiences a two-dimensional in-plane periodicity and is confined in the third dimensions by various light guiding mechanisms~\cite{johnson1999guided}.  Our work shows that non-Hermitian point-gap topology can be explored in photonic crystals being a technological relevant platform, without the need of relying on systems described by the tight-binding model.


\let\oldaddcontentsline\addcontentsline
\renewcommand{\addcontentsline}[3]{}

\begin{acknowledgements}
This work is supported by a Vannevar Bush Faculty Fellowship from the U. S. Department of Defense (Grant No. N00014-17-1-3030). JZ acknowledges support from the Fulbright Future Scholarship.
\end{acknowledgements}

\appendix*
\section{Reciprocity leads to symmetric band structures}
\label{appendix}
The photonic band structure for any photonic crystals consisting of periodic dielectric variations can be determined using the following master equation derived from Maxwell's equations~\cite{joannopoulos2011photonic}:
\begin{equation}
 \nabla \times\left(\varepsilon^{-1} \nabla \times  \mathbf{H}_{\mathbf{k} } \right)=\left(\frac{\omega(\mathbf{k})}{c}\right)^{2}  \mathbf{H}_{\mathbf{k} }\:.
\label{eq_master}
\end{equation}
where $\mathbf{H}_{\mathbf{k} }$ is the magnetic field at a wavevector $\mathbf{k} $, $\omega(\mathbf{k})$ is the eigenfrequency, $\varepsilon$ is the permittivity tensor and $c$ is the speed of light. For a reciprocal system $\varepsilon(r) = \varepsilon^T(r).$ We shall use Bloch's theorem for the magnetic field 
\begin{equation}
\mathbf{H}_{\mathbf{k}}=e^{-i\mathbf{k}\cdot \mathbf{r}}u_\mathbf{k}(r).
\end{equation}
With this, Eq.~\eqref{eq_master} becomes
\begin{equation}
\underbrace{(-i \mathbf{k}+\nabla) \times\varepsilon^{-1}(-i \mathbf{k}+\nabla) \times }_{\mathrm{A}_{\mathbf{k}}}u_{\mathbf{k}}=\left(\frac{\omega(\mathbf{k})}{c}\right)^2 u_{\mathbf{k}},
\label{eq_newmaster}
\end{equation}
where $\mathrm{A}_{\mathbf{k}}$ is an operator. \\

We now prove that 
\begin{equation}
\omega(\mathbf{k}) = \omega(-\mathbf{k})
\label{eq_proooove}
\end{equation}
provided that $\varepsilon$ is a symmetric tensor, i.e.
$\varepsilon = \varepsilon^T.$ For this proof,  we use the fact that $\mathrm{A}_{\mathbf{k}}$ and its transpose $\mathrm{A}_{\mathbf{k}}^T$ have the same spectrum since for any operator $\mathrm{A}$
\begin{equation}
\operatorname{det}(\lambda \mathrm{I}-\mathrm{A})=\operatorname{det}\left(\lambda \mathrm{I}-\mathrm{A}^{T}\right),
\end{equation}
where $\mathrm{I}$ is the identity matrix. Thus to prove Eq.~\eqref{eq_proooove} it is sufficient to prove 
\begin{equation}
\mathrm{A}_{\mathbf{k}}=\mathrm{A}_{-\mathbf{k}}^{T}
\label{Ak_akt}
\end{equation}
for reciprocal systems.\\

To prove Eq.~\eqref{Ak_akt}, we notice that 
\begin{equation}
(\mathbf{k} \times)^{T}=-(\mathbf{k} \times)
\end{equation}
as can be seen using the vector identity $\mathbf{C} \cdot(\mathbf{k} \times \mathbf{D})=-(\mathbf{k} \times \mathbf{C}) \cdot \mathbf{D}$ where $\mathbf{C}$ and $\mathbf{D}$ are arbitrary 3-vectors. We also note that 
\begin{equation}
(\nabla \times)^{T}=\nabla \times
\end{equation}
as can be seen from $\nabla \cdot(\mathrm{C} \times \mathrm{D})=(\nabla \times \mathrm{C}) \cdot \mathrm{D}-(\nabla \times \mathrm{D}) \cdot \mathrm{C}$, where $\mathrm{C}$ and $\mathrm{D}$ are vector fields. Therefore,
\begin{align}
\mathrm{A}_{\mathbf{k}}^T &= (-i \mathbf{k}\times+\nabla \times)^{T}\left(\varepsilon^{-1}\right)^{T}(-i \mathbf{k}\times+\nabla \times)^{T} \\
&= (i \mathbf{k}+\nabla)\times \varepsilon^{-1}(i \mathbf{k}+\nabla )\times \\
&= \mathrm{A}_{\mathbf{-k}}.
\end{align}
which proves Eq.~\eqref{eq_proooove}. For one dimension, Eq.~\eqref{eq_proooove} is equivalent to $\omega(k)=\omega(-k)$ and for two dimensions, it is equivalent to $\omega\left(k_{x}, k_{y}\right)=\omega\left(-k_{x},-k_{y}\right)$.
\nocite{apsrev41Control}
\bibliographystyle{apsrev4}
\bibliography{titleon,bib}

\let\addcontentsline\oldaddcontentsline

\newpage

\setcounter{figure}{0}
\setcounter{section}{0}
\setcounter{equation}{0}
\renewcommand{\thefigure}{S\arabic{figure}}
\renewcommand{\thesection}{S\Roman{section}}
\renewcommand{\thesection}{S\arabic{section}}
\renewcommand{\theequation}{S\arabic{equation}}

%

\end{document}